\documentclass[twocolumn,showpacs,preprintnumbers,amsmath,amssymb,superscriptaddress]{revtex4}
\usepackage{graphicx}
\usepackage{dcolumn}
\usepackage{bm}

\begin{document}


\title{Relevance of multiple-quasiparticle tunneling between edge states at $\nu=p/(2np+1)$}
\author{D. Ferraro}
\affiliation{%
Dipartimento di Fisica \& INFN, Universit\`a di Genova,Via Dodecaneso 33, 16146, Genova, Italy
}
\author{A. Braggio}
\affiliation{Dipartimento di Fisica \& LAMIA-INFM-CNR, Universit\`a di Genova, Via Dodecaneso 33,
16146, Genova, Italy}
\author{M. Merlo}%
\affiliation{Dipartimento di Fisica \& LAMIA-INFM-CNR, Universit\`a di Genova, Via Dodecaneso 33,
16146, Genova, Italy}
\author{N. Magnoli}
\affiliation{%
Dipartimento di Fisica \& INFN, Universit\`a di Genova,Via Dodecaneso 33, 16146, Genova, Italy
}
\author{M. Sassetti}ù
\affiliation{Dipartimento di Fisica \& LAMIA-INFM-CNR, Universit\`a di Genova, Via Dodecaneso 33,
16146, Genova, Italy}

\date{\today}
\begin{abstract}
  We present an explanation for the anomalous behavior in tunneling
  conductance and noise through a point contact between edge states in the
  Jain series $\nu=p/(2np+1)$, for extremely weak-backscattering and
  low temperatures [Y.C. Chung, M. Heiblum, and V. Umansky, Phys. Rev.
  Lett.  {\bf{91}}, 216804 (2003)]. We consider edge states with
  neutral modes propagating at finite velocity, and we show that the activation of their
  dynamics causes the unexpected change in the temperature power-law of
  the conductance. Even more importantly, we demonstrate that
  multiple-quasiparticles tunneling at low energies becomes the most relevant
  process. This result will be used to explain
  the experimental data on current noise where tunneling particles
  have a charge that can reach $p$ times the single quasiparticle charge.  In this paper
  we analyze the conductance and the shot noise to substantiate
  quantitatively the proposed scenario.


\end{abstract}
%
 \maketitle
 \textit{Introduction.}  Noise experiments in point contacts have been
 crucial to demonstrate the existence of fractionally charged
 quasiparticles (qp) in Fractional Quantum Hall
 systems~\cite{Laughlin83}. In particular, it was proved that for
 filling factor $\nu=p/(2np+1)$, with $n,p\in\mathbb{N}$, the qp
 charge is given by $e^*=e/(2np+1)$
 \cite{dePicciotto97,Saminadayar97,Reznikov99}.  A suitable framework
 for the description of these phenomena is provided by the theory of
 edge states \cite{Wen90,Wen91}. For the Laughlin series ($p=1$) a
 chiral Luttinger Liquid theory ($\chi$LL) with a single mode was
 proposed and shot-noise signatures of fractional charge were
 devised \cite{Kane94}.  For the Jain series~\cite{Jain89}
 ($p\ge 1$), extensions were introduced by considering $p-1$ additional
 hierarchical fields, propagating with finite velocity~\cite{Wen92},
 or two fields, one charged and one topological and
 neutral~\cite{Lopez99,Chamon07}.  At intermediate temperatures, the
 experimental observations of tunneling through a point contact with
 $\nu=1/3$~\cite{Roddaro04} are well described by the exact solution
 of the $\chi$LL theory~\cite{Fendley95} which interpolates between
 the strong- and the weak-backscattering limits.  However, at low
 temperatures and weak backscattering the current presents unexpected
 behaviors \cite{Heiblum03,Roddaro03,Roddaro04}.  For instance, the
 backscattering conductance decreases for $T\to 0$ instead of
 increasing as the theories would require.  This discrepancy was
 recently investigated and different mechanisms of renormalization of
 tunneling exponents were proposed to account for it: coupling with
 additional phonon modes \cite{Rosenow02}, interaction effects
 \cite{Papa04,Mandal02} or edge reconstruction~\cite{Yang03}.  For
 filling factors with $p>1$, there are other intriguing transport
 experiments on a point contact at low temperature and extremely weak
 backscattering~\cite{Heiblum03,Heiblum03PE} which, to our knowledge,
 are not yet completely understood.  The main puzzling observations
 for $\nu=2/5$ and $\nu=3/7$ are: i) a change in the power-law scaling
 of the backscattering current with temperature; ii) an effective
 tunneling charge, as measured with noise, that can reach the value $p
 e^*$ for ultra-low temperatures $T<20$ mK.

 In this Letter we propose a unified explanation of the above open
 points.  We will describe infinite edges with two fields, one charged
 and one neutral, following the Lopez-Fradkin theory~\cite{Lopez99,Chamon07}.
 However, differently from that approach, where the neutral mode is
 non-propagating and guarantees only the appropriate fractional statistics
 of qp excitations, we assume a {\em finite} velocity of propagation, smaller than the charged mode
velocity.
We will show that
 the energy scaling of the single qp tunneling is modified by the
 dynamics of neutral modes~\cite{Lee98}. This will be sufficient to
 explain a change in slope of the linear conductance vs $T$.  However, in
 order to find an ``effective'' tunneling charge larger than $e^*$ at very low
 temperature it is necessary to demonstrate that tunneling is
 dominated by an agglomerate of qps. We will show that in the presence
 of a {\em finite} bandwidth of the neutral mode this is indeed the
 case.

 \textit{Multiple-qp processes}. We start to describe tunneling through a point contact
 in a Hall bar with right/left edges ($j=R/L$) of infinite length
 \cite{Chamon07, Lopez99}.  Edge $j$ consists of a charged mode
 $\phi^c_j$ and a neutral mode $\phi^n_j$, mutually
 commuting.  The commutation
 relations of the fields are $[\phi_j^{c/n}(x),\phi_j^{c/n}(x')]=i \pi
 \eta^{c/n} \nu_{c/n} \textrm{sgn}(x-x')$, with $\eta^{c/n}=+/-$,
 $\nu_c=\nu$ and $\nu_n=1$.  Electron number density is given by the
 relation $\rho_j(x)=\partial_x \phi_j^c(x)/2\pi$. The
 real-time action $\mathcal{S}_j$ is ($\hbar=1)$
\begin{eqnarray}
\label{eq:action}
\mathcal{S}_j=&&\frac{1}{4\pi\nu}\int\!\! dt dx\  \partial_x \phi_j^c
(-\partial_t-v_c\partial_x)\phi_j^c+\nonumber \\
&&\frac{1}{4\pi}\int\!\! dt dx\  \partial_x \phi_j^n (+\partial_t-v_n\partial_x)\phi_j^n,
\end{eqnarray}
where the neutral mode $\phi^n_j$ is counterpropagating with respect
to $\phi^c_j$ and has a velocity $v_n\ll v_c$ \cite{footnote}.
Consequently, the relation between the bandwidths
$\omega_n=v_n/a$ and $\omega_c=v_c/a$ will be $\omega_{n}\ll\omega_c$, where $a^{-1}$ is the momentum
cut-off.

The operator that annihilates an agglomerate of $m$ qps for the $j$-th edge can be written
 in the bosonized form
\begin{equation}
\label{eq:noperator}
\Psi^{(m)}_j(x)=\frac{\mathcal{F}^m}{\sqrt{2\pi a}}e^{i [\sqrt{\alpha_m}
\phi_j^c(x)+\sqrt{\beta_m}\phi_j^n(x)]}\,.
\end{equation}
Here, $\mathcal{F}^m$ corresponds to the ladder operator for changing
the number of $m$-qps. It plays the role of a Klein factor, and in
lowest order in tunneling can be neglected.  The coefficients are determined by
requiring that $\Psi^{(m)}_j(x)$ satisfies the appropriate commutation
relations with the electron density
$[\rho_j(x),\Psi_j^{(m)}(x')]=-m(\nu/p)\delta(x-x')\Psi_j^{(m)}(x')$, and the statistical
properties $\Psi^{(m)}_j(x) \Psi^{(m)}_j(x')= \Psi^{(m)}_j(x') \Psi^{(m)}_j(x)
e^{-i \theta_m \textrm{sgn}(x-x')}$. The statistical angle is~\cite{Wilczek90}
\begin{equation}
\label{eq:theta}
\theta_m=\pi m^2\left(\frac{\nu}{p^2}-\frac{1}{p}-1\right)+2\pi k,
\end{equation}
where $k\in\mathbb{Z}$ takes into account the $2\pi$ periodicity.
Thus, for every value of $p$, one has
\begin{equation}
\label{eq:beta}
\alpha_m=\frac{m^2}{p^2}\,;\quad
\beta_m=m^2 \left(1+\frac{1}{p}\right)-2 k\,.
\end{equation}
Eq.~(\ref{eq:beta}) admits several solutions labelled by different $k
\leq k^{\rm max}$ with $k^{\rm max}=\mathrm{Int}[m^2(1+1/p)/2]$ where
$\mathrm{Int}[x]$ is the integer part of $x$.  For a given
$m$ there is a family of operators $\Psi^{(m)}_j$ with the
same fractional properties but, as we will see, different scaling
behavior. The local scaling dimension $\Delta_m$ of the $m$-agglomerate
operator is defined as half the power-law exponent at long times
($|\tau|\gg 1/\omega_c,1/\omega_n$)~\cite{Kane92} in  the two-point
imaginary time Green function $\mathcal{G}_m(\tau)=\langle
T_\tau[\Psi^{(m)}_j(0,\tau)\Psi^{(m)\
  \dagger}_j(0,0)]\rangle\propto\tau^{-2 \Delta_m}$.  At $T=0$ the
Green function is
\begin{equation}
\label{eq:Gret}
\mathcal{G}_m(\tau)=\frac{1}{2\pi a}\left(\frac{1}{1+\omega_c |\tau|}\right)^{g_c
\nu \alpha_m}\left(\frac{1}{1+\omega_n |\tau|}\right)^{g_n\beta_m},
\end{equation}
where one can clearly recognize in the last term the dynamical
contribution of the neutral modes.  The scaling dimension is then
$\Delta_m=(g_c\nu \alpha_m +g_n\beta_m)/2$.  Note that in order to
take into account possible additional interaction effects we
considered in Eq.(\ref{eq:Gret}) renormalization parameters
$g_{c,n}\geq 1$.
They correspond to the renormalization of the dynamical exponents induced by a coupling of the fields
with independent dissipative baths~\cite{Rosenow02}. The microscopic models underlying these
renormalizations were extensively treated in
literature~\cite{Rosenow02,Papa04,Mandal02,Yang03} and will not be
specifically discussed here. Note that the renormalizations do not
affect the statistical properties of the fields, which depend only on
the equal-time commutation relations, i.e.  the field algebra.
The \textit{most relevant} operator in the $m$-family will then have the
minimal value $\Delta_m^{\rm min} =[g_c\nu \alpha_m +g_n\beta^{\rm
  min}_m]/2$ given by the minimal value of  $\beta_m$ in Eq.~(\ref{eq:beta})
\begin{equation}
\label{eq:beta1}
\beta_m^{\rm min}=m^2(1+1/p)-2k^{\rm max}\,.
\end{equation}
Let us now identify the dominant process for specific cases.  In the
Laughlin series ($p=1$) one finds $\beta^{\rm min}_m=0$, and therefore
the single-qp tunneling ($m=1$) is always the dominant one since
$\Delta^{\rm min}_m=m^2\Delta^{\rm min}_1$.  A different scenario is
present for $p\geq 2$. Here one has for $m=1$ $\beta^{\rm
  min}_{1}=1+1/p$, while for the $p$-agglomerate $\beta^{\rm
  min}_{p}=0$.  This allows to conclude that agglomerates with $m> p$
are never dominant: $\Delta^{\rm min}_{m>p}>\Delta^{\rm min}_{m=p}$.

To find the most relevant operator one has to choose within the class
with $1\leq m\leq p$. In the bare case, $g_{n,c}=1$, one can show that
the $p$-agglomerate is the most relevant for $p\leq 6$.  With
renormalized exponents $g_{n,c}>1$ the analysis is still possible but
more cumbersome; we limit here the discussion to $p=2,3$, which
are directly connected with the experiments at
$\nu=2/5,3/7$~\cite{Heiblum03}.  It is furthermore possible to show with the above
relations that the $p$-agglomerate is always dominant in the parameter
region $g_n/g_c>\nu (1-1/p)$, while otherwise the single qp tunneling
prevails.

We conclude by emphasizing that, for a non-propagating neutral mode
with $v_n=\omega_n=0$, the single-qp processes will {\em always}
dominate because the neutral mode does not contribute to the scaling.

\textit{Transport}. In this part we restrict the analysis of tunneling
through the point contact at $x=0$ to $\nu=2/5$ and $\nu=3/7$. In
these cases we consider the two most dominant processes only: the
single qp and the agglomerate of $p$ qps.  The tunneling Hamiltonian is
$H_T=t_1 \Psi^{(1)\ \dagger}_{R}(0) \Psi^{(1)}_{L}(0)+t_p \Psi^{(p)\
  \dagger}_{R}(0)\Psi^{(p)}_{L}(0)+\textit{h.c.}$ with $t_1$ and $t_p$
the tunneling amplitudes.  As already discussed, here the operators
$\Psi_j^{(m)}$ are the most relevant representatives in the $m$-family.  In the
weak-backscattering limit the tunneling rates in lowest order are
($m=1,p$ and $k_B=1$)
\begin{equation}
\label{eq:rate}
\Gamma_m(E)=\gamma_m\!\int_{-\infty}^{+\infty}\!\!\!\!\!dt\  e^{i Et}\  e^{-[\alpha_m W^c(t)
+\beta_m^{\rm min} W^n(t)]}\,,
\end{equation}
with $\gamma_m= (|t_m|/2\pi a)^2$, and $W^{c/n}(t)=\sum_{j}\langle
[\phi_j^{c/n}(0,0)-\phi_j^{c/n}(0,t)]\phi_j^{c/n}(0,0)\rangle$ the
time-dependent bosonic correlation functions.  The explicit expression
of the kernel is $W^{r}(t)=g_{r} \nu_r\ln[(1+i\omega_r
t)\mathbf{\Gamma}(\eta_r)^2/|\mathbf{\Gamma}(\eta_r+i T t)|^2]$ where
$\eta_r=1+T/\omega_r$ with $r=c,n$ and $\mathbf{\Gamma}(x)$ is the Gamma
function~\cite{Weiss99}.  In the following, we assume that the neutral
mode bandwidth $\omega_n$ can be comparable with $T$ and with the external
voltage energy $e^* V$, while the charge bandwidth $\omega_c$ is taken
as the largest cut-off energy.
\begin{figure}
\begin{center}
\includegraphics[width=0.47\textwidth,clip=true]{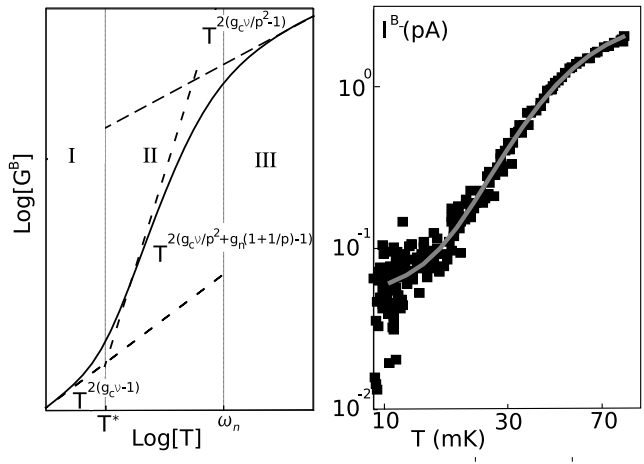}
\vskip-1cm
\caption{a) Sketch of the backscattering conductance $G^B$ vs
  temperature in a log-log plot.  The dashed lines are the
  asymptotic power laws and the solid line is the conductance
  in different temperature regimes: I low, II intermediate, and III high $T$. In this
  scheme the parameters are chosen with $T^*\ll\omega_n$,
  $\omega_n\ll\omega_c$, and $g_c\nu/p^2>1$. b) Comparison between the
  theoretical backscattering current $I^B$ (solid gray line) and the
  experimental data (black squares) at $\nu=2/5$ ($p=2$) from
  Ref.~\cite{Heiblum03} with courtesy of M. Heiblum.  Plotting parameters:
  $g_c=3$, $g_n=4$, $\omega_n=50$ mK, $\omega_n/\omega_c=10^{-2}$,
  $T^*=20$ mK, $e^*V=1.16$ mK, $\gamma_1/\gamma_2=1.66$ and $\gamma_1/\omega_c^2=4\cdot10^{-2}$.}\label{fig:1}
\end{center}
\end{figure}

We start now by comparing our theory with the experimental data. In lowest order, the total backscattering current
through the point contact is given by the sum of the
two independent processes contributions $I^B_{1}$ and $I^B_{p}$
\begin{equation}
I^B=\sum_{m=1,p}I^B_m= e^*\sum_{m=1,p}m (1-e^{-
  E_{m}/T})\Gamma_m(E_{m})\,,
\label{eq:current}
\end{equation}
with $E_m=m e^* V$ the energy for $m$-qp tunneling in the presence of
the bias $V$.  The linear backscattering conductance is then
$G^B(T)=\sum_{m=1,p}G^B_m(T)$ where $G^B_m(T)=(m e^*)^2\Gamma_m(0)/T$.
It will contribute to the total conductance via the relation $G(T)=\nu
e^2/2\pi -G^B(T)$. Before analyzing it numerically we discuss
qualitatively the different scaling regimes.  Let us start with
$G_1^B(T)$: for $T\ll\omega_n$ the neutral modes participate in the
temperature scaling giving $T^{2(g_c\nu\alpha_1+g_n\beta_1^{\rm min}
  -1)}$, while in the opposite limit $T\gg\omega_n$ the scaling is
driven by the charged modes only giving
$T^{2[g_c\nu\alpha_1-1]}$.
On the other hand,
the $p$-agglomerate follows the power law $G_p^B(T)\propto T^{2(
  g_c\nu\alpha_p-1)}$ with a scaling driven always by the charged
modes because $\beta^{\rm min}_{p}=0$. The total backscattering conductance
will depend on the relative weights between the single qp and
$p$-agglomerate contributions. We fix the ratio of the tunneling
amplitudes $t_{1}/t_p$ by introducing the temperature $T^*$ at which
$G_1^B(T^*)=G_p^B(T^*)$.

The experimental observations suggest the
relevance of the $p$-agglomerate at extremely low temperature so $T^* <\omega_n$
and the renormalization coefficients satisfy $g_n/g_c>\nu (1-1/p)$.

In this case the behavior of the backscattering conductance $G^B(T)$ presents
three distinct power-laws
\begin{equation}
\label{asymptotic}
G^B(T) \approx \begin{cases}
T^{2[\nu g_c-1]} &  T\ll T^*\\
T^{2[\nu g_c/p^2+g_n(1+1/p)-1]} & T^*\ll T\ll\omega_n\,,\\
T^{2[\nu g_c/p^2-1]} & \omega_n\ll T\\
\end{cases}
\end{equation}
where we explicitly used Eqs.~(\ref{eq:beta}) and (\ref{eq:beta1}).
A sketch of these behaviors is shown in Fig.\ref{fig:1}a in a log-log
plot. The solid line is the backscattering conductance and the dashed lines are
the three different asymptotic power laws in Eq.(\ref{asymptotic}).
At very low temperatures (region I) the
$p$-agglomerate dominates, while at higher temperatures (region II
and III) the single qp is dominant. Note that the intermediate
temperatures regime (II), where the neutral modes are effective,
will be accessible only if $T^*\lll\omega_n$. Otherwise, we expect a
mixing of region II and I.  Fig.\ref{fig:1}b shows the
backscattering current $I^B$ in Eq.~(\ref{eq:current}) for $\nu=2/5$ (solid gray line) evaluated
numerically. The parameters were
adjusted in order to fit the experimental data, black squares, taken
from Fig.2a of Ref.~\cite{Heiblum03}. With respect to the sketch in
Fig.\ref{fig:1}a, the best fit of the experimental data is mainly given by region II, where the
$p$-agglomerate is not fully effective.
We warn however the reader that due to the restricted experimental
range of temperatures (roughly one decade) it is not possible to
extract meaningful values for power-law exponents.
Anyway, from the fitting  an estimate of the neutral modes bandwidth
of $\omega_n\sim 50$ mK appears reasonable.  This fact could explain why in
several experiments, performed at higher temperatures, the effects of
the neutral modes are not easily detectable.

\textit{Shot-noise.} As shown in the experiments, direct information
concerning the effective charge transferred through the point
contact can be unambiguously obtained via the current noise spectrum $S$ at zero frequency.
In the following we analyze the shot-noise regime with $T\ll e^*V$.
For weak backscattering the different tunneling processes are
independent and the transport through the point contact has a
Poissonian nature~\cite{Kane94,Fendley95}.  The total noise is then
the sum of the two backscattering currents with proportionality factors given by the
corresponding tunneling charges, namely $e^*$ for single qp tunneling, and $p e^*$ for the $p$-agglomerate,
$S\approx 2 e^*(I_1^B+pI_p^B)$. Then the effective charge $q_{\rm eff}$ of the
tunneling process will be evaluated from the behavior of  the Fano factor $F=S/2 e I^B$,
via the relation $q_{\rm eff}=e F$.

For simplicity we consider the limit $T=0$.
The current (\ref{eq:current}) can be evaluated
without any further assumption
\begin{eqnarray}
\label{eq:rateT0}
I^B_m&=&m  \frac{4e^*\pi\gamma_m}{\omega_c^{a_m}\omega_n^{b_m}}\frac{e^{-E_m/\omega_c}}{
\mathbf{\Gamma}(a_m+b_m)}E_m^{a_m+b_m-1}\nonumber\\&\times
&\!\!_1F_1\left(b_m,a_m+ b_m,\frac{E_m}{\omega_c}-\frac{E_m}{\omega_n}\right)\,,
\end{eqnarray}
with $_1F_1(a,b,z)$ the Kummer confluent hypergeometric function,
$a_m=2g_c \nu \alpha_m$ and $b_m=2 g_n \beta_m$.  Similarly to the
conductance temperature scaling, the current exhibits different
regimes.  For $E_1\gg \omega_n$, the single qp contribution scales as $I^B_1\propto E_1^{2g_c \nu/p^2-1}$, while for
$E_1\ll\omega_n$ it receives additional contributions from the neutral
modes $I^B_1\propto E_1^{2[g_c \nu/p^2+g_n(1+1/p)]-1}$.  This twofold
power law is present only for the single qp tunneling since the
$p$-agglomerate current $I_p$ depends only on the charged mode dynamics
$I^B_p\propto E_p^{2\nu g_c-1}$. We define $V^*$ as
the voltage at which the two current contributions are equal,
$I^B_1(V^*)=I^B_p(V^*)$.  From the previous scaling argument we conclude that
for $V\ll V^*$  the $p$-agglomerate dominates, while for $V\gg V^*$  single qp tunneling is more relevant.

\begin{figure}
\begin{center}
\includegraphics[width=.48\textwidth]{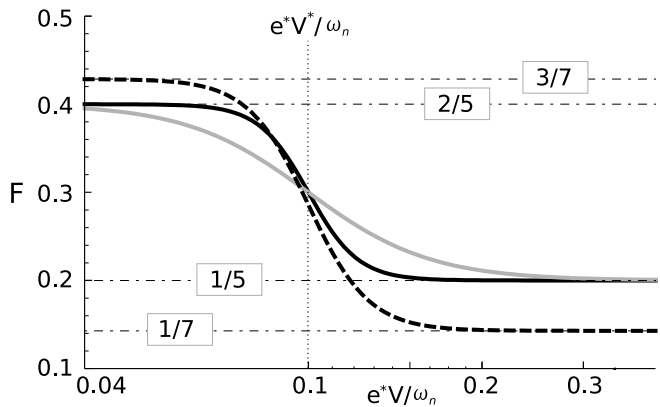}
\vskip-0cm
\caption{Zero temperature Fano factor $F$ vs source-drain potential
  $e^* V/\omega_n$ in log scale. The solid lines represent $\nu=2/5$
  for $g_c=3$, $g_n=2$, $\kappa=2.75$ (gray), and $g_c=3$, $g_n=4$ $\kappa=5.25$ (black).
  The dashed line is
  $\nu=3/7$ with $g_c=3$, $g_n=4$ and $\kappa=4.26$. Other parameters $\omega_n/\omega_c=10^{-2}$,
$e^*V^*/\omega_n=0.1$.}
\label{fig:2}
\end{center}
\end{figure}

In Fig.~\ref{fig:2} the Fano factor is shown as a function
of the external voltage for $\nu=2/5$ (solid) and $\nu=3/7$ (dashed). One can easily recognize two
regimes with distinct effective charges: for $V\gg V^*$ the noise is
dominated by the single-qp processes and  $q_{\rm eff}=e^*$, while
for $V\ll V^*$ the $p$-agglomerate will prevail with $q_{\rm eff}=p e^*=\nu e$.
Note that the width of the transition region is determined by the
difference between the power-law exponents of $I_1$ and $I_p$.
Indeed, defining the ratio $\kappa=\Delta_1^{\rm min}/\Delta_p^{\rm
  min}>1$, one has a sharper transition for larger $\kappa$ values.
This is confirmed by the behavior in the figure for $\nu=2/5$ with different values of
$\kappa$  obtained by changing the ratio $g_n/g_c$ (see caption).  The
smoothness of the Fano factor could be then relevant to determine the
renormalized parameters and the voltage at which the
$p$-agglomerate tunneling is clearly visible.

We observe that the above results on the possibility to detect an
effective tunneling charge $q_{\rm eff}=pe^*$ will remain valid also
at finite temperatures as long as $T\ll e^*V^*$. At higher
temperature the dominance of the $p$-agglomerate is progressively
compromised.

The above facts, i.e. the smoothness of the transition and/or the temperature effects,
could explain why in the experiment for $\nu=3/7$ the limiting value
$F=3/7$ is not fully reached in the experimental window while for $\nu=2/5$ the limiting value
is observed.

\textit{Conclusion}. We have shown that $p$-qp agglomerates can be the most dominant
tunneling process through a point contact at extremely low
temperatures in the weak-backscattering regime.  Direct signatures of
this relevance are shown  in the behavior of the shot noise.
The main point underlying this result is the assumption of {\em physical} neutral modes
propagating at finite velocity. Their dynamical activation
affects the single-qp tunneling scaling and makes it less relevant than multiple-qp tunneling. In addition,
we explain the double power law observed in the temperature scaling of the
backscattering current.

Though in this work we mainly investigated the experimental observations of
Ref.~\cite{Heiblum03}, we expect that the results
could be also relevant for other experimental situations.
A new generation of experimental studies of shot noise in point contacts at extremely low
temperatures are desirable in order to shed light on the intriguing physics of fractional qp agglomerates.

\textit{Acknowledgments}. We thank M. Heiblum for useful discussions. Financial support by the EU via Contract
No. MCRTN-CT2003-504574 is gratefully acknowledged.


\begin{thebibliography}{24}
\bibitem{footnote}{The mode
  counterpropagation is required by the signature of the Lopez-Fradkin
  K matrix.}
\end{thebibliography}
\end{document}